\documentclass[conference]{IEEEtran}
\IEEEoverridecommandlockouts
\usepackage{amsmath,amsfonts}
\usepackage{algorithmic}
\usepackage{algorithm}
\usepackage{array}
\usepackage[caption=false,font=normalsize,labelfont=sf,textfont=sf]{subfig}
\usepackage{textcomp}
\usepackage{stfloats}
\usepackage{url}
\usepackage{verbatim}
\usepackage[dvipdfmx]{graphicx}
\usepackage{cite}

	\usepackage{color}
	\usepackage{latexsym}
	\usepackage{multirow} 
	
	\def\epsfsize#1#2{\ifnum#1>\hsize\hsize\else#1\fi}
	
	\newif\ifdraft
	\draftfalse 
	
	\ifdraft
	\newcommand{\controne}[1]{\color{red} #1 \color{black}}
	\newcommand{\contrtwo}[1]{\color{green} #1 \color{black}}
	\newcommand{\contrthree}[1]{\color{blue} #1 \color{black}}
	\else
	\newcommand{\controne}[1]{#1}
	\newcommand{\contrtwo}[1]{#1}
	\newcommand{\contrthree}[1]{#1}
	\fi

\begin{document}

\title{Model-based Development for Autonomous Driving Software Considering Parallelization}

\author{\IEEEauthorblockN{Kenshin Obi}
\IEEEauthorblockA{\textit{Graduate School of}\\
\textit{Science and Engineering}\\
\textit{Saitama University}}
\and
\IEEEauthorblockN{Takumi Onozawa}
\IEEEauthorblockA{\textit{Graduate School of}\\
\textit{Science and Engineering}\\
\textit{Saitama University}}
\and
\\
\IEEEauthorblockN{Hiroshi Fujimoto}
\IEEEauthorblockA{\textit{Software Division}\\
\textit{eSOL Co., Ltd}}
\and
\IEEEauthorblockN{Takuya Azumi}
\IEEEauthorblockA{\textit{Graduate School of}\\
\textit{Science and Engineering}\\
\textit{Saitama University}}
}

\maketitle
\IEEEpubid{\makebox[\columnwidth]{© 2024 IEEE. Personal use of this material is permitted.
DOI: 10.1109/ETFA61755.2024.10711120\hfill}}
\IEEEpubidadjcol
\vspace*{1.5\baselineskip} 

\begin{abstract}
In recent years, autonomous vehicles have attracted attention as one of the solutions to various social problems. However, autonomous driving software requires real-time performance as it considers a variety of functions and complex environments. Therefore, this paper proposes a parallelization method for autonomous driving software using the \textit{Model-Based Development}~(MBD) process. The proposed method extends the existing \textit{Model-Based Parallelizer}~(MBP) method to facilitate the implementation of complex processing. As a result, execution time was reduced. The evaluation results demonstrate that the proposed method is suitable for the development of autonomous driving software, particularly in achieving real-time performance.
\end{abstract}

\begin{IEEEkeywords}
Model-based development, multi-core processor, autonomous driving software
\end{IEEEkeywords}

\section{Introduction}
Autonomous driving systems have been attracting attention in recent years as the population ages and rural areas depopulate. These social problems have led to reductions in unprofitable public transportation systems. The concentration of stores and large hospitals has expanded the living areas of the seniors and increased the amount of time driving by the seniors. Older drivers are at higher risk of accidents than other generations, due to a decline in cognitive skills and exercise capacity~\cite{Older_drivers}. In addition, truck drivers have become increasingly important as online shopping has become more prevalent since COVID-19~\cite{COVID-19}. Autonomous driving systems are expected to be used for public transportation and logistics.

On the other hand, building autonomous driving systems has been a difficult task. One of the reasons for the difficulties in constructing the system is the pursuit of real-time performance~\cite{Autoware}. Autonomous vehicles need to complete the process before the deadline to avoid traffic accidents. Simply simplifying the processing and increasing the processing speed will reduce recognition accuracy, making it difficult to ensure safety. In addition, the software must be tested in a variety of situations to ensure safety. However, conducting all of these tests on actual vehicles would be difficult from both a cost and reproducibility standpoint. Furthermore, the work required to go back to the design phase to solve problems found in these tests is also a major burden on development.

\textit{Model-Based Development}~(MBD)~\cite{ROS_Self-Driving}, which is used in in-vehicle software development, can be a solution to this problem~\cite{Autoware_Benchmark}. Simulation of MBD can be used to verify the operation of autonomous driving software in scenarios that are difficult to reproduce in a real vehicle. However, MBD alone does not improve the real-time performance of autonomous driving software. The use of multi/many cores is a method to improve the real-time performance of autonomous driving software~\cite{many-core}. \textit{Model-Based Parallelizer}~(MBP)~\cite{MBP} is a method that can automatically generate parallelized code optimized for the operating environment from a model developed using MBD. Therefore, MBP reduces the time required to assign tasks to each core and more easily reduces the execution time of autonomous driving software.

This study proposes a developing method for designing autonomous driving software using MBD and generating parallelized code. Existing MBP has the problem that the blocks usable in task parallel are limited. In Simulink, add-ons called Toolbox allow processes that require many blocks and time to design to be expressed in a small number of blocks. However, existing MBP does not support Toolbox blocks. In addition, a solution to the problem of decreasing the number of blocks due to the use of Toolbox blocks and code descriptions is presented. The proposed method performs task parallelization of the model by utilizing Toolbox. Furthermore, a design method that suppresses the decrease in parallelism due to the decrease in the number of blocks is presented. The contributions of this research are as follows:
\begin{itemize}
    \controne{\item Reduction of execution time by model-based parallelization for models with Toolbox.}
    \contrtwo{\item Improvement of development efficiency by model-based parallelization for models with Toolbox.}
    \contrthree{\item Reduction of execution time in sensor and control models for autonomous driving software utilizing multi-core processors.}
\end{itemize}

The paper is organized as follows. The system model is described in Section~\ref{section:system model}. The details of the proposed method are pursued in Section~\ref{section:design}. An evaluation of the proposed method is presented in Section~\ref{section:evaluation}. Comparisons with related studies are made in Section~\ref{section:related work}, and conclusions and discussions are given in Section~\ref{section:conclusion}.

\IEEEpubidadjcol
\section{System Model}\label{section:system model}
\begin{figure}[tbp]
	\centerline{\includegraphics[width=0.9\linewidth]{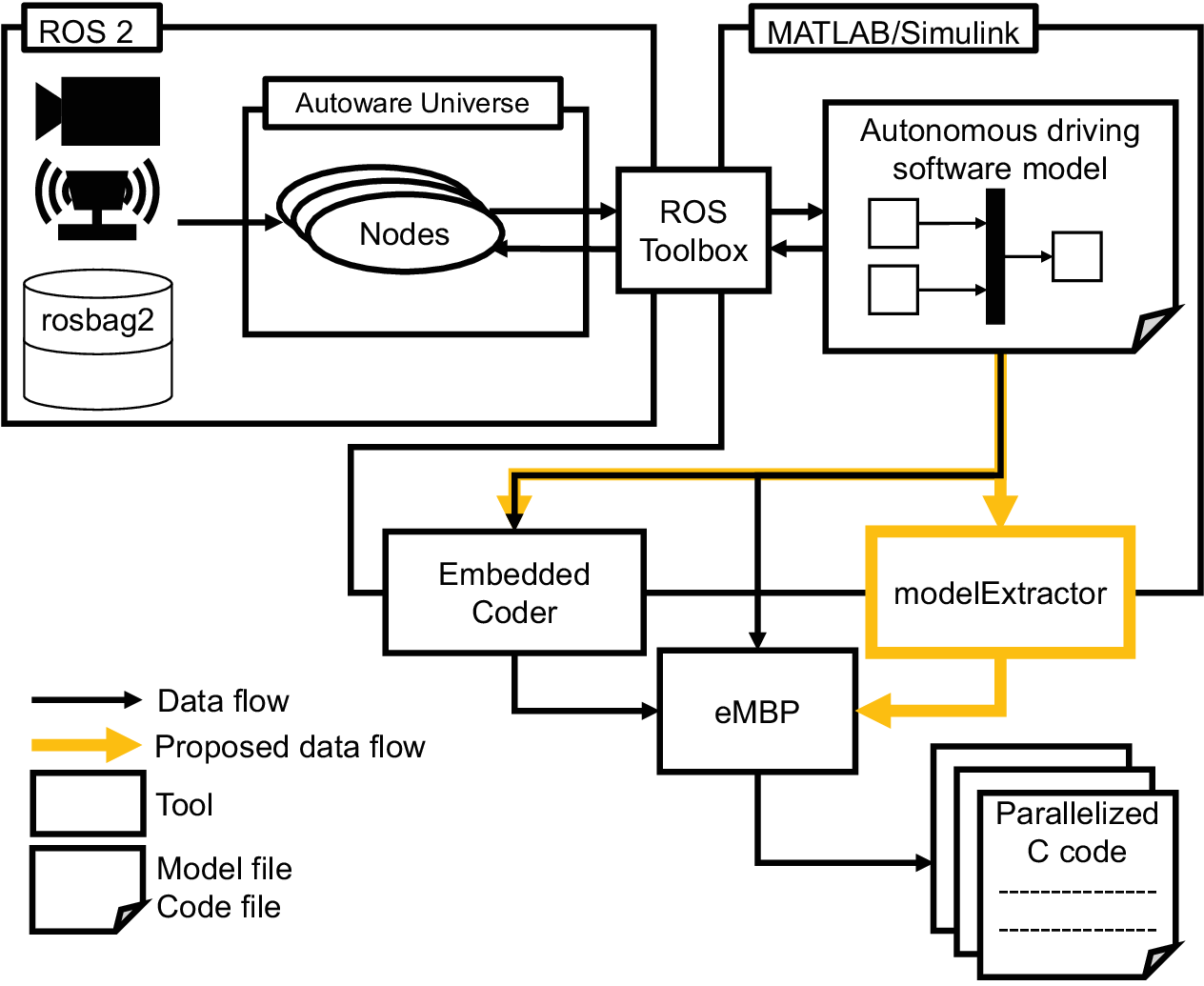}}
	\caption{System model.}
	\label{System model}
\end{figure}
A method for generating parallelized code from a model of autonomous driving software utilizing Toolbox is proposed in this study. The system model is shown in Fig.~\ref{System model}. The input of the proposed method is an autonomous driving software model, and the output is a parallelized C code. The remainder of this section describes each element in the system model.

\subsection{ROS~2}\label{subsection:ROS}
\begin{figure}[tbp]
	\centerline{\includegraphics[width=0.8\linewidth]{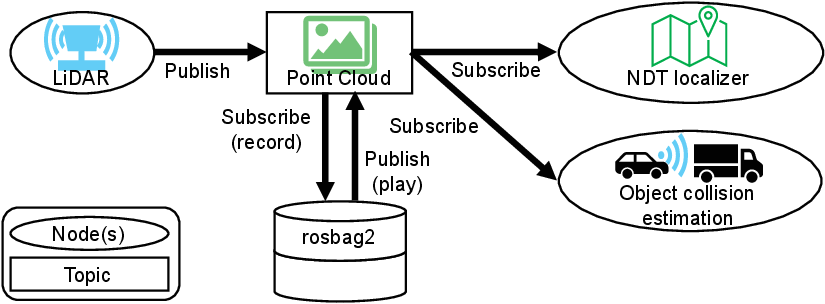}}
	\caption{Publish-subscribe model.}
	\label{Pub/Sub model}
\end{figure}

ROS~2~\cite{ROS2} is a middleware suite used in robot development. ROS~2 includes libraries and tools for hardware abstraction, device drivers, and message communication. ROS~2 software has the advantage of high reusability to divide the processing into units called nodes. Data communication between nodes is performed using a method called the publish-subscribe model with topics, as shown in Fig.~\ref{Pub/Sub model}. 
In this communication scheme, sending a topic is referred to as publishing, and the publishing node sends a message to a topic without assuming the receiving node. Receiving a message from a topic is referred to as subscribing, and the subscribing node specifies the type and topic name of the topic to be received. ROS cannot be used in commercial embedded systems conventionally due to single point of failure, insufficient real-time performance, unstable operation, and high-power consumption. Thus, ROS~2, which is optimized for embedded real-time systems, e.g., autonomous driving systems, was developed to address these issues.

ROS~2 changes the communication protocol to DDS and improves operational stability by eliminating single points of bility. Thus, ROS~2 can now be used in commercial embedded systems, e.g., autonomous driving systems.

In this study, a node-by-node model of autonomous driving software was created, focusing on the ROS~2 node. This allowed the model to be validated by replacing the processing of autonomous driving software that has not been created with existing ROS~2-based autonomous driving software. Developing ROS~2-based autonomous driving software will further facilitate testing. ROS~2 has a topic storage functionality called rosbag2. By testing multiple times on the same topic, the output for the same input can be verified to be unchanged.

\subsection{Autoware Universe}\label{subsection:Autoware}
\begin{figure}[tbp]
	\centerline{\includegraphics[width=0.8\linewidth]{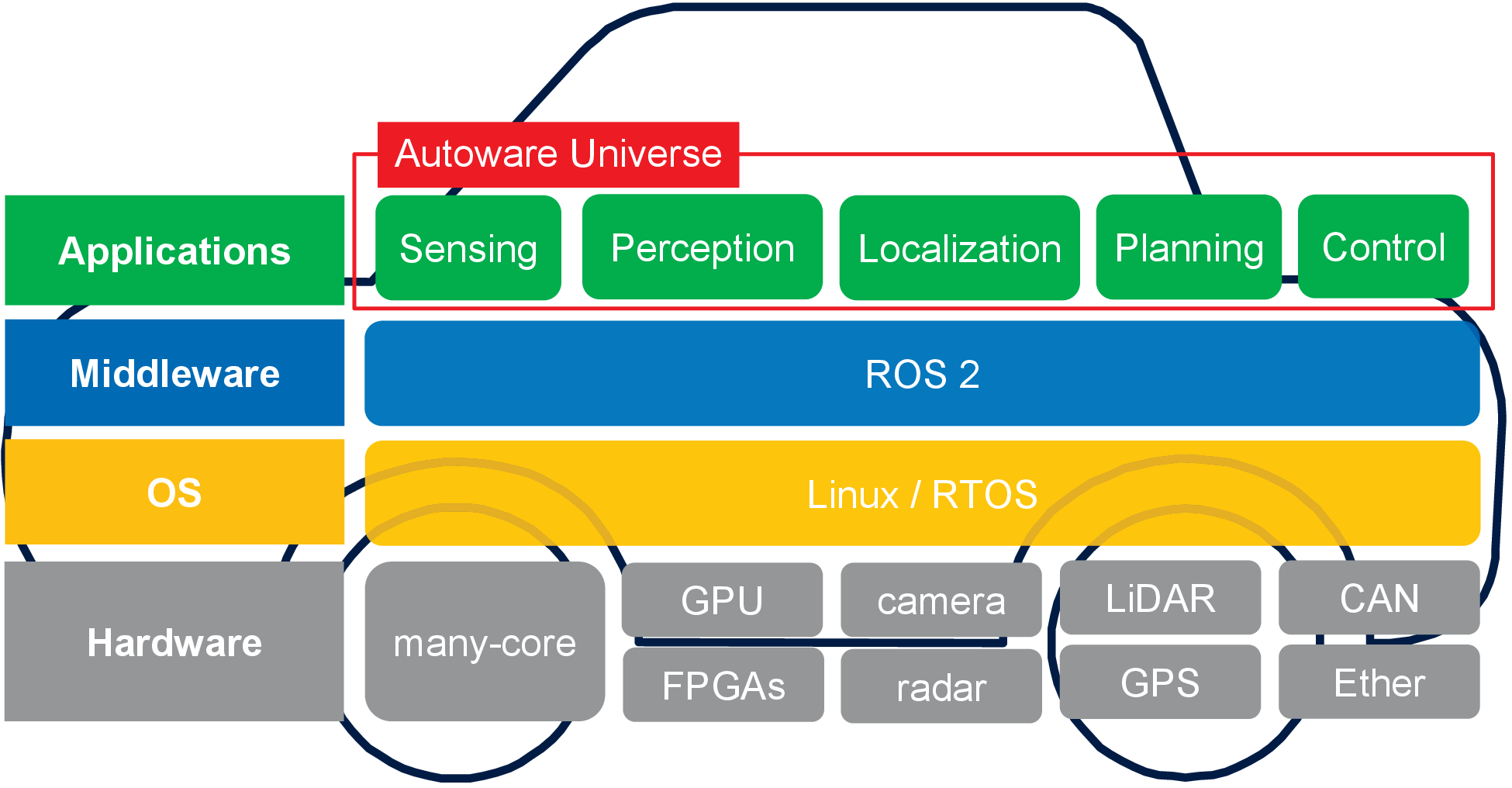}}
	\caption{Autoware functionality and operating environment.}
	\label{Autoware}
\end{figure}

The latest version of Autoware~\cite{AutowareOverview}, Autoware Universe~\cite{AutowareUniverse}, is open-source autonomous driving software using ROS~2. Using ROS~2, software can augment real-time capability and make effective decisions quickly. For example, the early recognition of near-miss cases and timely response will facilitate the avoidance of traffic accidents. As shown in Fig.~\ref{Autoware}, Autoware provides five autonomous driving software functionalities, e.g., sensing, localization, environment perception, path planning, and contol.

Note that Autoware is implemented as sequential code. Thus, parallelization requires rewriting the code, which incurs high time costs to ensure the identity of the operation. In addition, even if parallelization is performed once, the system needs to run on different hardware configurations for open-source software. In such cases, optimal operation cannot be expected unless parallelization is performed again. Therefore, MBD is suitable for the development of autonomous driving software, as it can be developed based on parallelization without limiting the operating platform.

\subsection{MATLAB/Simulink}\label{subsection:MBD}
MATLAB/Simulink, provided by MathWorks~\cite{MathWorks}, is the de facto standard software for MBD. MATLAB is a programming and numerical computing platform that is used for data and algorithmic analysis. MATLAB is particularly suitable for three types of processes such as matrix calculations, numerical analysis, and image processing. Here, the programs are written in the MATLAB language. In addition, Simulink is development software that expresses processing in units of blocks and connects these blocks for programming. blocks and connects these blocks for programming. Moreover, the use of subsystems, which are groups of multiple blocks, allows processing to be hierarchically organized and improves readability and reusability. Thus, the flow of the system can be visualized clearly using this software. Furthermore, the blocks that constitute a Simulink model are written in the MATLAB language, and add-on functionalities are available to extend the functionality of MATLAB/Simulink. These functionalities allow for efficient and effective ROS~2-based software development and embedded software development in MATLAB/Simulink.

ROS Toolbox add-on enables Simulink to handle ROS or ROS~2 systems. With ROS Toolbox, topics are represented in Simulink as \textit{bus} structures, where the \textit{bus} is a type of signal line in Simulink that combines data of different types into a single data structure. A hierarchical structure can also be created by giving the \textit{bus} the input of the block that creates the \textit{bus}. In addition, this add-on allows developers to express a publish-subscribe model in a single block in an effective manner. Note that part of topic does not store data directly and requires conversion of values for processing. For example, the coordinates of each point in a point cloud topic are represented as real values, but these are stored as unsigned integer values within the topic. ROS Toolbox also includes a block that is responsible for such a transformation. Thus, this add-on allows ROS~2 developers to use Simulink.

Embedded Coder add-on provides optimized code generation for embedded devices. MATLAB/Simulink each provides its own code generator. However, they are not constrained by requirements such as memory usage and execution speed. Embedded Coder has information on a variety of embedded processors and microcontrollers, such as ARM, Intel SoCs, and Renesas microcontrollers, and can generate code optimized for these hardware. The generated code is optimized based on the scheduler’s control and priority constraint requirements.

\begin{figure}[tbp]
	\centerline{\includegraphics[width=1.0\linewidth]{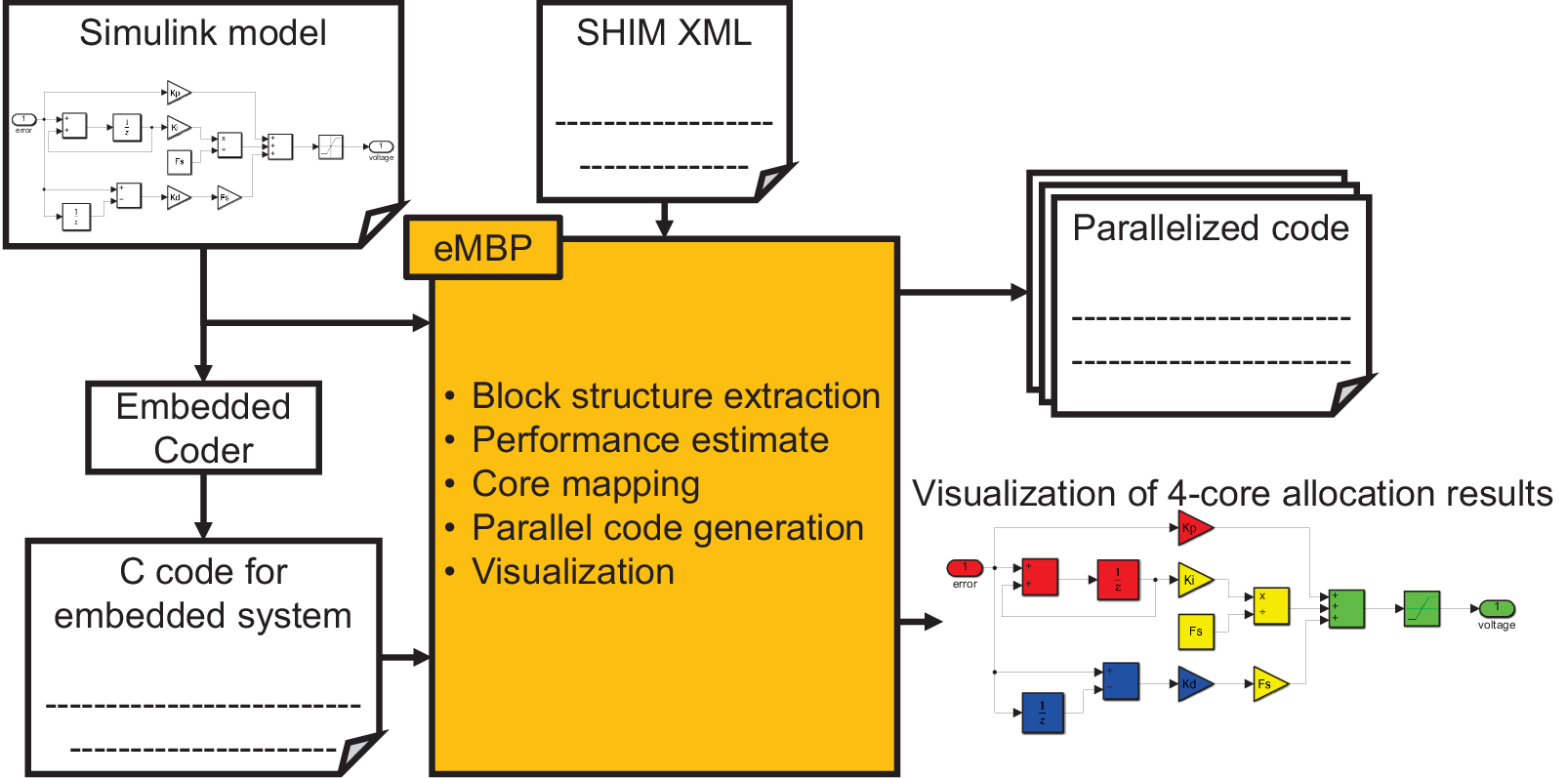}}
	\caption{eMBP input/output structure.}
	\label{MBP}
\end{figure}

\subsection{eMBP}\label{subsection:MBP}
eMBP is a tool, provided by eSOL Corporation, used to generate parallelization code from Simulink models~\cite{Mapping2022}. As shown in Fig.~\ref{MBP}, the input to eMBP is the Simulink model, C code, and SHIM. First, the model is split into blocks, the code fragments corresponding to the blocks are combined with information about the block structure, and the result is exported to BLXML in XML format. Subsequent processing is performed by adding information to BLXML. The information to be added is performance information and core allocation information. Here, the performance information is calculated using the inputs to Software-Hardware Interface for Multi-Many-Core (SHIM)~\cite{SHIM}, a common interface that abstracts hardware characteristics. SHIM describes performance information such as the number of cores, core performance, and the time required for inter-core communication. Therefore, eMBP can utilize this information to estimate execution time and optimize core allocation in the operating environment. Finally, eMBP automatically generates parallelized code with task parallelism in block units based on the information added to BLXML. In addition, the core allocation status of the parallelization results can be visualized.

\section{Proposed Method}\label{section:design}
\begin{figure}[tbp]
	\centerline{\includegraphics[width=1.0\linewidth]{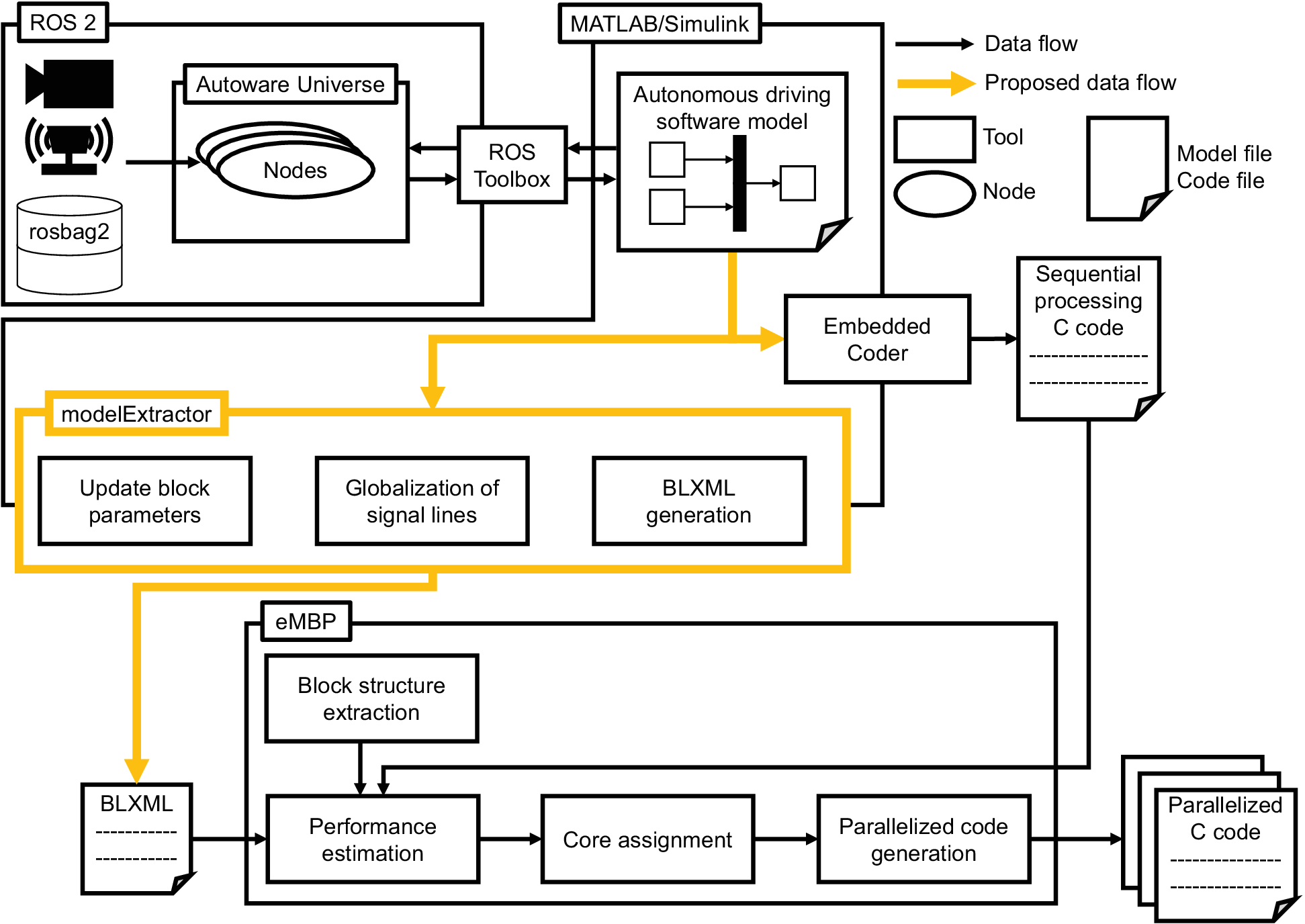}}
	\caption{Proposed method.}
	\label{Proposed method}
\end{figure}
A parallelization method for ROS~2-based autonomous driving software using MBP approach is proposed, as shown in Fig.~\ref{Proposed method}. Software parallelization can be divided into two main approaches: task parallelization and data parallelization. Task parallelization is a method in which software is divided into multiple tasks, and each task is executed on a different processor. At this time, the tasks to be divided must have no dependencies. Therefore, the filtering process that extracts the input data is difficult to divide in task parallelism. Data parallelization is a method in which input data is divided into multiple parts, and each part is processed by a different processor. The proposed method aims to improve real-time performance by performing task parallelism and utilizing the existing data parallelism method~\cite{data_parallel}.

\subsection{Task Parallelism for Models with Toolbox}\label{subsection:task_p}
\controne{
Simulink models can generate task-parallelized code using eMBP. Toolbox in 
Simulink models can generate task-parallelized code using eMBP. Toolbox in MATLAB/Simulink has functions that implement complex calculations such as sensing. Therefore, the use of Toolbox is important for improving the development efficiency of autonomous driving software. However, block structure information, which is important for task parallelization in eMBP, cannot be generated from Toolbox blocks. This is because eMBP targets only the basic blocks of Simulink. Therefore, the proposed method extracts block structure information from models with Toolbox to enable task parallelization in eMBP. Task parallelization with eMBP is performed in four steps, as shown in Fig.~\ref{Proposed method}. In the first step, BLXML is generated by extracting block structure information from the Simulink model. In the second step, the execution time is estimated to use as a reference for core allocation, and the core allocation is performed in the third step. Based on these results, the final step generates parallelized code optimized for a specified number of cores or less. Among these steps, only the first step is affected by Toolbox.

The proposed method, modelExtractor, first updates the information of blocks and signal lines in the model. Instead of eMBP, modelExtractor extracts the block structure from the Simulink model. Therefore, modelExtractor must generate XML and instruct code generation according to the BLXML description format. To achieve this goal, two information updates are performed. The first is to make signal lines global variables and to make block and signal line names unique. eMBP treats the data handled by all signal lines as global variables and gives each line a unique name. In treating signal lines as global variables, modelExtractor makes block names unique in the model. At this time, it uses block paths to take advantage of the unique names given in the model design. The parameters of the signal line are then updated to treat the signal line as a global variable. The name of the signal line should be a combination of the source block and port number.

The second is to update the parameters of the bus signal, which handles multiple signal lines. The Bus Selector block, which selects a bus signal element, specifies the name of the signal line when it is combined with the bus signal when selecting a signal line. Since the signal line names have changed with the globalization of signal lines, an error occurs when trying to select a signal line that does not exist as it is, and code generation cannot be performed. Therefore, the parameters of the Bus Selector block are referenced to correct the information. First, the structure of the input signal lines and the positions of the necessary elements from the output signal lines are recorded before making the signal lines global variables. Next, the Bus Selector block is read to determine the structure of the input signal line after it has been made a global variable. Finally, it reselects the elements of the output signal line based on the recorded element positions and the read input signal line structure.

modelExtractor supports Masked Subsystem blocks in addition. A mask is a Simulink interface that hides the contents of a block. It has the advantage of simplifying the appearance of Simulink models by encapsulating blocks. On the other hand, the disadvantage is that the contents of Masked Subsystem blocks can only be viewed on the GUI of Simulink. Therefore, block structure information could not be obtained. In this research, we found a parameter to check the contents of the Masked Subsystem block from the CUI. This allows block structure information to be obtained with modelExtractor.
}

\begin{figure}[tbp]
	\centerline{\includegraphics[width=\linewidth]{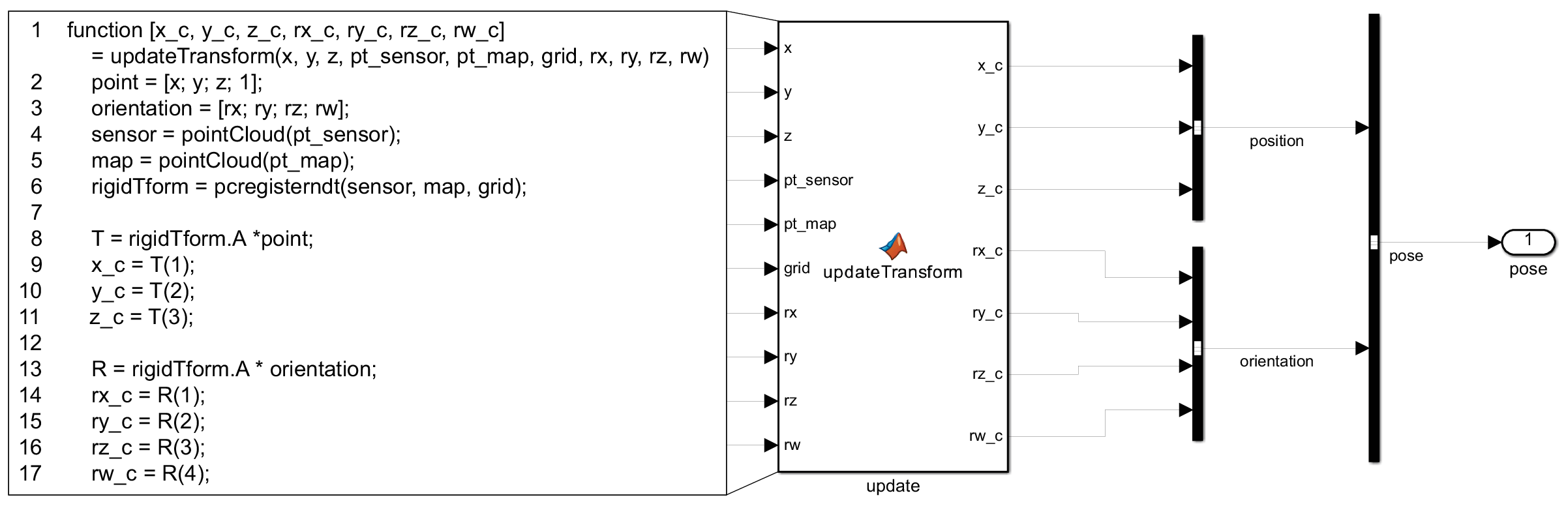}}
	\caption{Model implemented by sequentially writing MATLAB code.}
	\label{MC_c}
\end{figure}
\begin{figure}[tbp]
	\centerline{\includegraphics[width=\linewidth]{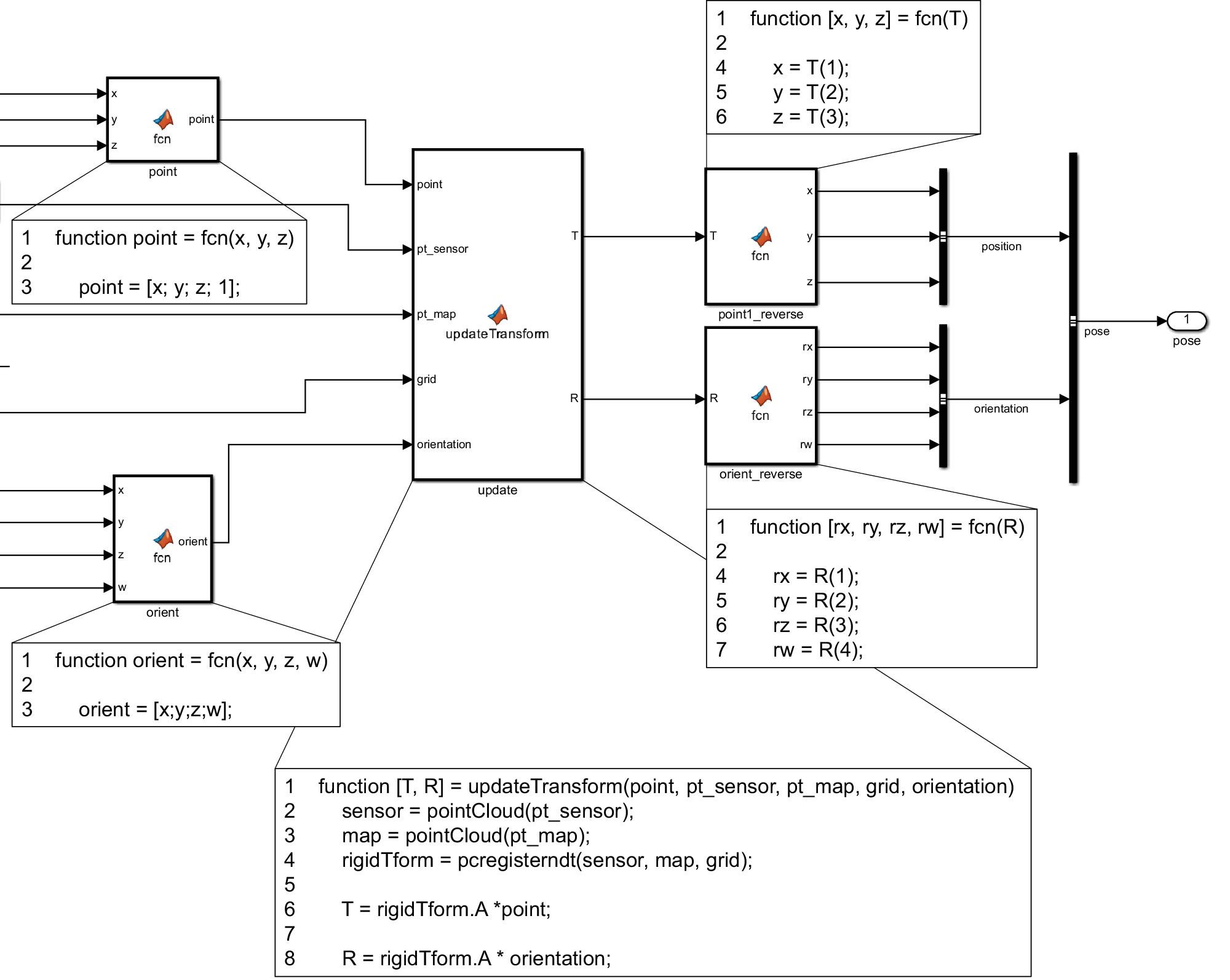}}
	\caption{Model implemented by splitting MATLAB code.}
	\label{MC_d}
\end{figure}

\subsection{Considerations in Model Design}\label{subsection:model_design}
\contrtwo{
Toolbox, an add-on to MATLAB/Simulink, streamlines the representation of autonomous driving software in a model. While Toolbox blocks simplify the implementation of complex formulas, decreasing the number of blocks reduces the number of cores allocated. This is due to the fact that task parallelism in eMBP operates at the block level, regardless of the unchanged computational load. The cause of this problem is the structure of the model, which is difficult to solve with modelExtractor and eMBP. The reason is that these tools refer to the structure, but do not provide editing functionalities. Consequently, designing a model that considers this limitation is crucial for optimizing execution time. In this subsection, two cases of model design innovations are described.

When using MATLAB code, the code should be divided into multiple blocks whenever possible. Simulink also allows development using MATLAB code, although certain Toolbox functions are not in Simulink blocks. These functions can be incorporated by adding a MATLAB Function block into the model and writing code within this block. In the basic design of eMBP without considering task parallelism, continuous code is used to create matrices and to edit a portion of matrix elements as modeled in Fig.~\ref{MC_c}. However, in the proposed method, the sequential description of the code reduces the number of blocks and cores to be allocated. Therefore, a method to prepare a different MATLAB Function block for each function, similar to the model illustrated in Fig.~\ref{MC_d}, was considered. However, Simulink has limitations on the data that can be handled by signal lines between blocks. In particular, data types classified as ``System objects,'' which encompass information on specific functions known as ``object functions,'' along with structural data, are incompatible with Simulink signal lines. In scenarios excluding such cases, the functions to be executed sequentially can be written in multiple MATLAB Function blocks. This increases the number of blocks in the model, thereby enabling eMBP to assign tasks across a greater number of cores.
}

\controne{
The disadvantages of using Toolbox can be compensated for by devising the model design. This allows the user to solve the problem of complex functionalities implemented in Toolbox becoming a bottleneck in execution time, while still reaping the benefits of reduced development time. In addition, dividing complex processing into multiple blocks facilitates allocating processing to a larger number of cores in eMBP parallelization. The division of the code and blocks reduces execution time bias among cores.
}

\section{Evaluation}\label{section:evaluation}
In this study, a method for generating parallelized code from Simulink models using both data-parallel and task-parallel approaches is proposed. The proposed method is evaluated in terms of the behavior and execution time of the generated parallelized code. Nodes in Autoware Universe are modeled and used for evaluation.

Two patterns of evaluation methods were used depending on the situation. The first method is to acquire data in the laptop environment, as shown in Table~\ref{env_pc}. This method was used to verify the behavior of the model to be evaluated. In this method, input/output topics of nodes in Autoware Universe are first recorded in rosbag under the Ubuntu environment. Next, the obtained rosbag is replayed, and the model's output topics are recorded in the rosbag by simulating the model. In this way, the processing results for the same input data can be compared.

The second method is to run the code generated from the model under the embedded computer, as shown in Table~\ref{env_many}, to obtain the data. This method generates and compares three types of code for a single process. The three types of code are the sequential processing code generated by the standard Simulink code generator, the parallelized code generated by the existing method, and the parallelized code generated by the proposed method. The first method confirms that the model and the output are consistent with equivalent nodes in Autoware. Then, the second evaluation method was used.

\begin{table}[tbp]  
\caption{Evaluation environment for laptop computers}
\label{env_pc}
\footnotesize
\begin{center} 
{\tabcolsep=1mm
\begin{tabular}{|l|c|}
 \hline
  Processor&Intel Core-i7-10750H CPU @ 2.60 GHz\\
 \hline
  Number of Cores&6\\
 \hline
  Memory&32.0 GB\\
 \hline
  GPU&NVIDIA GeForce GTX 1650 Ti\\
 \hline
  OS&Ubuntu22.04 LTS\\
 \hline
  MATLAB/Simulink&R2023a\\
 \hline
  ROS 2 version&Humble\\
 \hline
  eMBP version&2.3.1\\
 \hline
\end{tabular}
}
\end{center}
\end{table}

\begin{table}[tbp]
\caption{Evaluation environment for embedded computers}
\label{env_many}
\footnotesize
\begin{center} 
{\tabcolsep=1mm
\begin{tabular}{|l|c|}
 \hline
  Platform&Raspberry Pi 5\\
 \hline
  Processor&Broadcom BCM2712~(Arm Cortex-A76$ \times 4$)\\
 \hline
  Number of Cores&4\\
 \hline
  Memory&8.0 GB\\
 \hline
  GPU&Broadcom VideoCore V\hspace{-1.2pt}I\hspace{-1.2pt}I\\
 \hline
  OS&Raspberry Pi OS bookworm\\
 \hline
\end{tabular}
}
\end{center}
\end{table}

\subsection{Model of Autonomous Driving Software Used for Evaluation}\label{subsection:ads_model}
Three different models representing the three types of Autoware nodes used in the evaluation are described. The name, type, and summary of processing for each of the modeled nodes are shown in Table~\ref{ads_model}. All models use ROS Toolbox as they require ROS~2 communication. The remainder of this subsection describes the details of each node's processing and the characteristics of the model.

\begin{table*}[tbp]  
\caption{Model of autonomous driving software} 
\label{ads_model}
\footnotesize
\begin{center} 
{\tabcolsep=1mm\begin{tabular}{|c|c|l|}
 \hline
  Name of node&Node type&Summary of processing\\
 \hline
  voxel\_grid\_downsample\_filter&sensing&\begin{tabular}{l}Divide the point cloud into a grid and extract one point from each grid.\end{tabular}\\
 \hline
  random\_downsample\_filter&sensing&\begin{tabular}{l}Extract the point cloud to have less than or equal to the specified maximum number of data.\end{tabular}\\
 \hline
  trajectory\_follower&control&\begin{tabular}{l}Calculate control values for accelerator, brake, and steering from current and future attitude and speed.\end{tabular}\\
 \hline
\end{tabular}
}
\end{center}
\end{table*}

\subsubsection{voxel\_grid\_downsample\_filter}\label{subsubsection:voxel}
voxel\_grid\_downsample\_filter divides the received PointCloud2 type point cloud data into a grid and creates point cloud data with only the representative points of each grid. This node further reduces the amount of computation by reducing the amount of data through the integration of neighboring points. In addition, the model at this node uses a function included in \textit{Computer Vision Toolbox} for point cloud extraction.

\subsubsection{random\_downsample\_filter}\label{subsubsection:random}
random\_downsample\_filter is a node that randomizes the number of points in the received PointCloud2 data to the number specified by the parameter. This node limits the maximum size of the point cloud data, making it easier to predict the execution time of subsequent processing.

\subsubsection{trajectory\_follower}\label{subsubsection:trajectory}
trajectory\_follower is a node that receives the future attitude, the future speed, and the current attitude to generate control values for the accelerator, brake, and steering. This node enables an autonomous vehicle to perform the three basic actions of a car: running, turning, and stopping. In the model representing this node, the process for generating control values in the vertical direction and horizontal direction are implemented as separate blocks. These blocks are implemented in \textit{Automated Driving Toolbox}, and the calculation of control values utilizes nonlinear control methods~\cite{Stanley}.

\subsection{Evaluation of Execution Time for Models Using Toolbox by Task Parallelism}\label{subsection:tp_toolbox_time_eval}
\controne{The effect of task parallelization on models with Toolbox was evaluated. For the evaluation, trajectory\_follower was used among the models described in Section~\ref{subsection:ads_model}. As in Section~\ref{subsection:dp_toolbox_time_eval}, the proposed method was used to generate BLXML with block structure information from the models with Toolbox. The generated BLXML could be allocated to multiple cores when task parallelization was performed using eMBP.
\begin{figure}[tbp]
	\centerline{\includegraphics[width=0.75\linewidth]{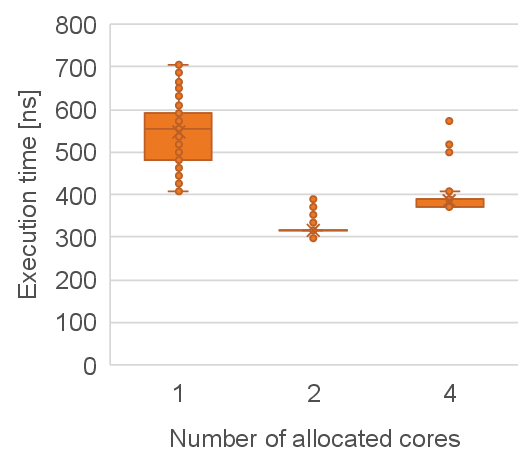}}
	\caption{Relationship between number of allocated cores and execution time for models with Toolbox.}
	\label{tp_time}
\end{figure}
Fig.~\ref{tp_time} shows the relationship between the execution time of the parallelized code generated by the proposed method and the allocated cores. A comparison of the execution time for each allocated core shows that the optimal allocation for this model is two cores. Two cores represent the optimal allocation for two reasons. First, the model can be executed in less than one~\textmu s even before parallelization. Therefore, the execution time that can be reduced by parallelization is short, and the increased communication overhead has a larger impact. The second is that the model itself is driven by two main paths. As described in Section~\ref{subsection:ads_model}, the main processing of the model is realized in separate blocks for vertical and horizontal control. In addition, each process is independent. Therefore, it is easy to execute each process on a separate core. Thus, the proposed method enables task parallelization even for models using Toolbox, which simplifies model design and reduces execution time.}

\subsection{Evaluation of Execution Time for Models Using Toolbox by Data Parallelism}\label{subsection:dp_toolbox_time_eval}
\controne{The effect of data parallelization on models with Toolbox was evaluated. For the evaluation, random\_downsample\_filter was used among the models described in Section~\ref{subsection:ads_model}. Using the proposed method and modelExtractor, BLXML with block structure information was generated from models with Toolbox. However, since the model uses ROS Toolbox, Embedded Coder generates C++ code, which breaks the restrictions of eMBP. Parallelization was attempted by eliminating ROS Toolbox blocks to handle the input/output. The ROS Toolbox block basically performs data input/output and type conversion, which has no effect on the parallelization results. As a result of the core allocation, eMBP allocates one core for processing of this model. However, although the processing of the Toolbox block can be data-parallel, it cannot be allocated to multiple cores for only one block. Therefore, the Toolbox block that can be data parallelized is divided into multiple blocks and allocated to multiple cores. As a result, the number of divisions matched the number of allocated cores. Then, the execution time of each block estimated by eMBP was checked. BLXML can be generated using the proposed method.
\begin{figure}[tbp]
	\centerline{\includegraphics[width=0.75\linewidth]{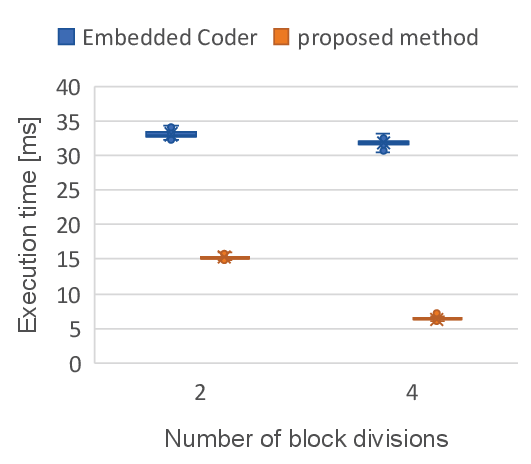}}
	\caption{Relationship between number of block divisions and execution time for models with Toolbox.}
	\label{dpt_time}
\end{figure}

The execution time of the code generated from the model designed using Toolbox is shown in Fig.~\ref{dpt_time}. As the measured execution time shows, the proposed method reduces the execution time compared to Embedded Coder, which generates sequential processing code. The fact that the proposed method takes less than half the time of the Embedded Coder-generated code can be attributed to further optimization by eliminating unnecessary processing. This is because eMBP splits the code generated by Embedded Coder into block units when generating parallel code and combines the pieces of code again. These results indicate that the proposed method can allocate processing to multiple cores with limitations, depending on the design of the model. Therefore, the proposed method can reduce the execution time.}

\subsection{Evaluation of Model Designs Using Toolbox Blocks}\label{subsection:tp_desgin_eval}
\contrtwo{The changes in terms of model design due to the use of Toolbox were evaluated. The number of lines of code and blocks to be added to realize the functionality used in the nodes of the autonomous driving software are shown in Table~\ref{line_and_block}.}
\begin{table*}[tbp]  
\caption{Model changes using Toolbox} 
\label{line_and_block}
\footnotesize
\begin{center} 
{\tabcolsep=1mm
\begin{tabular}{|c|c|c|c|}
 \hline
  Node&Implementation Method&Number of Lines of Code&Number of Blocks\\
 \hline
  &Autoware Universe&201&-\\
 \cline{2-4}
  \begin{tabular}{c} voxel\_grid\_downsample\_filter\end{tabular}&\begin{tabular}{c}MATLAB/Simulink without Toolbox\end{tabular}&54&68\\
 \cline{2-4}
  &\begin{tabular}{c}MATLAB/Simulink with Toolbox\end{tabular}&4&25\\
 \hline
  &Autoware Universe&164&-\\
 \cline{2-4}
  \begin{tabular}{c} random\_downsample\_filter\end{tabular}&\begin{tabular}{c}MATLAB/Simulink without Toolbox\end{tabular}&50&70\\
 \cline{2-4}
  &\begin{tabular}{c}MATLAB/Simulink with Toolbox\end{tabular}&6&27\\
 \hline

\end{tabular}
}
\end{center}
\end{table*}
\contrtwo{As shown in Table~\ref{line_and_block}, Toolbox reduces the number of lines of code and the number of times blocks are added. This is the effect of multiple blocks and multiple lines of code being combined into one block. Although the number of blocks is reduced to less than half, half of the reduced blocks are used for data input/output and data structure creation. Therefore, the impact of the large reduction in the number of blocks on the calculation cost is kept small compared to the reduction in the number of blocks. The number of items to check when verifying model operation is reduced. Thus, Toolbox reduces the amount of work in model design and simplifies the operation verification process, contributing to the efficient development of autonomous driving software. The reduced human resources can be used to improve the accuracy of functionalities and pursue the safety of autonomous driving software.}

\section{Related Work}\label{section:related work}
In this section, the proposed method is compared to eight previous studies on autonomous driving software or parallel code generation. First, comparisons are made with studies in terms of the combination of autonomous driving software and Simulink. Then, studies on simulation technology were discussed. Finally, studies on parallel code generation methods were introduced. The comparative studies on autonomous driving software and parallelization are organized point by point in Table~\ref{RW}.
\begin{table*}[tbp]  
\caption{Summary of existing studies on autonomous driving software and parallelization} 
\label{RW}
\footnotesize
\begin{center} 
{\tabcolsep=0.75mm
\begin{tabular}{|l|c|c|c|c|c|c|}
 \hline
  &Autonomous driving software&MATLAB/Simulink&Simulation&Benchmark&Task parallelization&ROS~2\\
 \hline
 IEEE ICMA 2020~\cite{ROS2_Autonomous_Driving_Platform} &\checkmark&\checkmark&&&&\checkmark\\
 \hline
 MDPI Applied Sciences 2021~\cite{PIL}  &\checkmark&\checkmark&&\checkmark&&\\
 \hline
 IEEE RSP 2019~\cite{Toolbox} &\checkmark&\checkmark&\checkmark&\checkmark&&\\
 \hline
 IEEE DESTION 2022~\cite{CPS} &\checkmark&\checkmark&\checkmark&&&\\
 \hline
 FISITA 2021~\cite{MIL}  &\checkmark&\checkmark&\checkmark&&&\\
 \hline
 Springer STTT 2019~\cite{PCR} &&&&&\checkmark&\\
 \hline
 IEEE/ACM PEHC 2021~\cite{OSCAR} &&&&&\checkmark&\\
 \hline
 This paper &\checkmark&\checkmark&\checkmark&\checkmark&\checkmark&\checkmark\\
 \hline
\end{tabular}
}
\end{center}
\end{table*}

\subsection{Combination of Autonomous Driving Software and Simulink}\label{subsection:Self-driving model}
Safety and reliability are primary factors for autonomous driving software. A High-Low Level Controller framework using ROS~2~\cite{ROS2_Autonomous_Driving_Platform} was proposed. In their study, they implemented a Simulink-compatible real-time \textit{Drive-by-Wire}~(DBW) kit with a \textit{controller area network}~(CAN) bus. This study shows that the combination of ROS~2 and the CAN bus made the proposed autonomous driving platform both safe and reliable.

The development of autonomous driving software must consider complex problems. A fully automated \textit{Processor-in-the-Loop}~(PIL) architecture~\cite{PIL} was demonstrated. This demonstration involved placing Autonomous vehicles in a racing competition. This experiment demonstrated that the PIL procedure can be used to test new features of autonomous vehicles quickly on a universal platform during the V-cycle.

An approach to developing autonomous driving software is to utilize ROS. Autoware Toolbox~\cite{Toolbox} was provided as a MATLAB/Simulink benchmark for Autoware, the ROS-based autonomous driving software. This facilitates the collaborative development of autonomous driving software with external organizations since it allows the use of the MBD approach common in the automotive industry.

Most systems for which MBD has been used in existing studies on autonomous driving software are vehicle control systems. In this study, the more computationally intensive sensing systems were also treated with models.

\subsection{Simulation Technology}\label{subsection:Simulation}
The \textit{cyber-physical systems}~(CPS) is a complex system that realizes its functionalities on a network. The behavior and characteristics of the CPS are analyzed via distributed simulation. The concerns associated with using multiple federations were clarified, and the time synchronization methods and data exchange between federations~\cite{CPS} were defined. As a result, the study assessed how to handle complex scenarios and identified a set of key mechanisms.

Model-in-Loop~(MIL) is a testing method that utilizes simulation technology to verify the control of a developed model. A virtual learning environment for control algorithms of autonomous driving software utilizing Toolbox in MATLAB/Simulink~\cite{MIL} was developed. As a result, vehicle control algorithms can be tested in the early development phase. In addition, more robust and accurate results were obtained in four different test scenarios.

Autonomous driving software is frequently verified by simulation, as verification in a real environment is often dangerous. In this study, Simulink was used to simulate the operation of the autonomous driving software and measure the execution time without using a real environment.

\subsection{Parallel Code Generation Methods}\label{subsection:paracode_gen}
The low cost of effective parallel software programming requires a high degree of expertise. A parallel programming pattern for a dataset called \textit{PCR}~\cite{PCR} was proposed. Using this pattern, platform-dependent parallelization programs can be generated automatically from platform-independent high-level specifications. As a result, the synthesized code can achieve performance that is comparable to low-level, unstructured platform-dependent programs.

Heterogeneous multicore processors consisting of a general-objective core and an accelerator core are widely used in various fields to achieve high performance and low power consumption. The \textit{OSCAR Parallelizing and Power Reducing compiler} and the \textit{OSCAR API}~\cite{OSCAR} were developed to generate parallel machine code automatically, considering optimal load balancing between cores and reduced data transfer overhead. The proposed compiler and API enable the task parallelization of programs, which results in faster execution and reduced power consumption.

Effective parallelization is necessary for applications that run in multicore environments. In this paper, parallelization using MBP is extended to be able to generate parallelization code from ROS~2 models.

\section{Conclusion}\label{section:conclusion}
This study proposes a method for generating parallelized code from models in order to reduce the execution time of autonomous driving software. The proposed method extends the existing MBP method to models with Toolbox, an add-on that makes it easy to implement complex functionalities. The proposed method was evaluated in terms of execution time on Raspberry~Pi~5 and model design by using models that represent different functionalities of autonomous driving software. 

The evaluation results show that the proposed method improves execution time and development efficiency. Furthermore, it is easier to analyze nodes that become bottlenecks in the pursuit of real-time performance. The future work is focused on establishing a multi-node core allocation method that takes into account the entire software. This is expected to improve real-time performance of complex systems.

\footnotesize
\bibliographystyle{IEEEtran}
\bibliography{bstcontrol,Reference}

\vfill

\end{document}